\documentstyle[12pt,epsfig]{article}
\author{Juli\'an Candia, Luis N. Epele and Esteban Roulet\\
{\it Phys. Dept., U. of La Plata, CC67, 1900, La Plata, Argentina.}}

\title{Cosmic ray photodisintegration and the
knee of the spectrum}

\begin{document}

\maketitle

\begin{abstract}
We explore in some detail the scenario proposed
to explain the observed knee of the cosmic ray (CR) spectrum as due to
the effects of photodisintegration of the CR nuclei by interactions
with optical and soft UV photons
in the source region. 
We show that the photon column densities needed
to explain the experimental data are significantly lower than
those obtained in previous estimations which neglected
multinucleon emission in the photodisintegration process.
We also treat more accurately the photodisintegration
thresholds, we discuss the effects of photopion production
processes and the neutron escape mechanism, identifying the
physical processes responsible for the qualitative features of
the results. This scenario would require the CR nuclei to
traverse column densities of $\sim 5 \times 10^{27}- 
2 \times 10^{28}$~eV/cm$^2$
after being accelerated in order to reproduce the observed knee,
and predicts that the CR composition should become lighter above
$\sim 10^{16}$~eV.
\end{abstract}

\section{Introduction}

It is widely accepted nowadays that cosmic rays (CRs) with energies
per particle up to about $10^{18}$~eV are protons and nuclei of
galactic origin. Furthermore, it is well established that the
full CR energy spectrum has a power-law behavior with
a steepening taking place at the so-called knee,
corresponding to an energy $E_{knee}=3\times 10^{15}$~eV.
Although it is well known that the CR composition below the knee
has a significant heavy component,
its behavior beyond the knee remains somewhat controversial.
Indeed, above $10^{14}$~eV/nucleus the information
about the CR mass composition has to be drawn from 
extensive air shower observations at ground level 
with the results from different experiments not 
always being compatible. Moreover, the theoretical
 predictions depend on the hadronic model adopted and this
can introduce further uncertainties on the inferred composition.
The model dependence in certain observables can however 
be useful to discriminate 
among the hadronic models using the experimental data, and in this 
comparison QGSJET and VENUS turn indeed to be favoured against 
other models such as SIBYLL \cite {fowler}.
Thus, the CR mass composition beyond the knee is not 
definitively established yet, with some observations
\cite {swordy} suggesting that it turns lighter and others
instead suggesting that the heavier components become
dominant \cite {amen1,ag00,ka01,ar00}.

The attempts made so far to explain the physical origin of the 
knee of the cosmic ray spectrum
can be roughly classified into three kind of models.
One of them exploits the possibility that the
acceleration mechanisms could be less effective above the
knee \cite{fich, joki, koba}, while a second one assumes 
that leakage from the Galaxy plays the
dominant role in suppressing CRs above the knee \cite{syro,wdow,ptus}.
The third scenario, originally proposed by Hillas \cite{hillas}, 
considers nuclear photodisintegration processes
(and proton energy losses by photomeson
production), in the presence of a background
of optical and soft UV photons in the source region,
as the main responsible for the change in the CR spectrum.
The first two scenarios are based on a rigidity dependent
effect which consequently 
produces a change in the spectral slope of each nuclear component with
charge $Z$ at an
energy $\sim E_{knee}Z$. Hence, it
removes first the protons
and the lighter nuclei and only at larger energies affects the heavier nuclei,
predicting that the CR mass composition should become 
heavier beyond the knee. On the contrary, 
the third scenario
predicts a lighter CR composition above the knee due
to the disintegration of the propagating nuclei.
The predictions of this last scenario have been 
worked out by Karakula and Tkaczyk \cite{karak} some years ago. 
The aim of the present paper is to re-evaluate in detail the
propagation of nuclei and protons in the source region
following their approach, but taking
into account some additional features that turn out to be
important for the determination of the source parameters 
required for the scenario to work.
In particular, we re-evaluate the nuclear photodisintegration rates
taking into account multinucleon emission processes, we
introduce more accurate threshold energies in the giant dipole
resonance for every stable nucleus (according to recent
remarks by Stecker and Salamon \cite{steck1}), and discuss the impact
of the neutron mechanism \cite{neutron} which would allow the escape
of neutrons from the source without further energy losses.
Moreover, we show that by introducing a non negligible lower cutoff to the
power-law photon distribution, the abrupt suppression
of the all-particle CR spectrum above $10^{17}$~eV
previously obtained \cite{karak} is prevented, and hence the
underlying reason for that effect is identified.
Let us also mention that although we focus our study here on
the scenario proposed to explain the observed steepening
at the knee, we expect that our general considerations about the treatment
of photodisintegration processes should also be relevant
in other contexts.

\section{The propagation of cosmic rays}

\subsection{The source and the photon background spectra}

As already mentioned, we deal with a model that considers that the CR 
nuclei are accelerated inside discrete sources which are surrounded 
by a very strong background of optical and soft UV
photons.
If the typical energy of the photons is in the optical range ($1-10$ eV),
the photodisintegration of CR nuclei will start to be efficient
at CR energies $E \geq A \times 10^{15}$~eV. This is so because in the 
CR rest frame the optical photon (which is boosted by a relativistic factor $\gamma \geq 10^6$) 
appears as an energetic gamma ray with $E \geq 1-10$~MeV,
i.e. capable of photodisintegrating the nucleus.
Hence, the basic assumption is that the CR spectrum emitted by the
source is a featureless extrapolation  of the spectra measured 
below the knee and that the observed change in slope is actually
 due to the CR interactions (i.e. the process
of nuclear photodisintegration and the photopion production
by protons) with the surrounding photon background.

We then assume that the source emits nuclei with mass number $A$
(with $A$ ranging from 1 to 56). Since there is only one stable
isotope for a nucleus of a given mass $A$ along most of the decay
chain from $^{56}$Fe to $^{1}$H, we take for definiteness a
unique charge $Z$ associated to
any given value of $A$. The differential fluxes emitted by the source
are assumed to be given by power-law distributions

\begin{equation}
\phi_i^0 (E) = \Phi_i^0  E^{-\gamma_{i}}
\end{equation} 

\noindent (where $i=56-A$ throughout).
The intensities $\Phi_i^0$ and spectral indices $\gamma_{i}$
were taken from
the detailed knowledge about CR mass composition below the
knee (for energies per particle above
$\sim  few \ \ Z \times 10^{10}$~eV) \cite{wiebel}.

For the photon background that surrounds the source we consider
that its spectrum follows either a (thermal) Planckian-type distribution
given by

\begin{equation}
n(\epsilon) = {\rho \over 2  \zeta (3)  (k_{B}T)^{3}}
\frac{\epsilon^{2}}{\exp(\epsilon/k_{B}T)-1}
\end{equation} 

\noindent where $k_{B}$ is the Boltzmann constant, $T$ the
absolute temperature and $\zeta$ the Riemann zeta function
(such that $\zeta (3) \approx 1.202$), 
or a power-law distribution given by

\begin{equation}
n(\epsilon) = \rho (\alpha-1)
\left(\frac{1}{\epsilon_{m}^{\alpha-1}}-\frac{1}{\epsilon_{M}^{\alpha-1}}\right)^{-1} \epsilon^{-\alpha}
\end{equation} 

\noindent where $\epsilon_{m}$, $\epsilon_{M}$ are the lower and upper
energy cutoffs and $\alpha$ the assumed spectral index.
For both distributions, $\rho$ represents the total number density and
$n(\epsilon)$ the differential number density corresponding to a photon energy 
$\epsilon$. Defining $\rho_E$ as the total energy density, one
has for the Planckian spectrum

\begin{equation}
\rho_E \approx 2.631  k_BT \rho \ \ , 
\end{equation}
while the analogous expression for the power-law spectrum is

\begin{equation}
\rho_E = {\alpha - 1 \over 2- \alpha } \left( \epsilon_M^{2-\alpha} -
\epsilon_m^{2-\alpha} \right) \left( {1 \over \epsilon_m^{\alpha - 1}} -
{1 \over \epsilon_M^{\alpha - 1}} \right)^{-1}   \rho \ \ .
\end{equation}
We adopt $\alpha =1.3$ as in ref. \cite{karak}, but
the introduction of a non negligible lower cutoff 
$\epsilon_{m}$ in the power-law spectrum excludes the presence
of an abundant number of low energy (infrared) photons that would
otherwise play the dominant role in the photodisintegration above 
$10^{17}$~eV, as was the case in ref. \cite{karak}. 
Regarding the spatial dependence of the radiation background  
we do not make any particular assumption, but the results will 
depend only on the integrated photon column density along 
the trajectory of the CR since its production at the source until it
leaves the region filled with radiation.

\subsection{Propagation of nuclei: photodisintegration}

The main mechanism of energy loss for nuclei with energy per particle
below $10^{18}$~eV propagating through these photon backgrounds is the
process of photodisintegration. In fact, a nucleus of mass $A=56-i$ and
Lorentz factor $\gamma=E/Am_pc^2$ that propagates through a photon distribution
$n(\epsilon)$ (as, for example, those of eqs. (2) and (3)) has a
probability of fission with  emission of $j$ nucleons given by 

\begin{equation}
R_{ij}(E)={1\over 2\gamma^2}\int_{\epsilon'_{thr,ij}/2\gamma}^\infty
{\rm d}\epsilon \ \
{n(\epsilon)\over \epsilon^{2}}\int_{\epsilon'_{thr,ij}}^{2\gamma\epsilon}
{\rm d}\epsilon'\epsilon'\sigma_{ij}(\epsilon'),
\end{equation} 
where $\sigma_{ij}$ is the corresponding photodisintegration cross
section, $\epsilon$ the photon energy in the observer's system and
$\epsilon'$ its energy in the rest frame of the nucleus. 
In order to calculate $R_{ij}$ we fitted $\sigma_{ij}$
with the parameters given in Tables I and II of ref. \cite{puget},
while the reaction thresholds $\epsilon'_{thr,ij}$
were taken from Table~I of ref. \cite{steck1}. A useful quantity to
take into account all reaction channels is the effective emission rate
given by

\begin{equation}
R_{i,eff}=\sum_{j\geq 1}jR_{ij} \ \ .
\end{equation} 

At low energies ($\epsilon'_{thr,ij} \leq \epsilon' \leq 30$ MeV),
the cross section $\sigma_{ij}(\epsilon')$ is dominated by the
giant dipole resonance and the photodisintegration proceeds chiefly
by the emission of one or two nucleons. Note that, according
to recent remarks pointed out in ref. \cite{steck1}, we employ
threshold energies that depend both on the nucleus and on the
number of emitted nucleons.
The improvement consists in shifting the
threshold energy, that was previously assumed
to be $\epsilon'_{thr}=2$~MeV for all reaction channels and 
all nuclei \cite{karak,puget}, to higher energies,
such that single-nucleon emission has now a typical threshold of
$ \sim  10$~MeV, while the double-nucleon emission energy threshold 
becomes typically $ \sim  20$~MeV.
Although the appearance of the knee in this scenario is ultimately
a threshold effect, 
the improved calculation
including the change in the specific threshold energies has however a minor
impact on the resulting emission rates when compared with
those obtained with a fixed threshold energy $\epsilon'_{thr} = 2$ MeV.
This can be appreciated in Figure 1, where
plots of $R_{i,eff}/\rho_{E}$ vs energy
per nucleon $E^*$ are shown for a $^{56}$Fe nucleus ($i=0$) propagating through
a thermal photon spectrum (with $k_{B}T=10$ eV). The emission rate $R_{0,eff}$
(calculated with the specific threshold energies) 
shows a small shift towards higher energies with respect to
that obtained by means of the fixed threshold energy (labeled 
$R_{0,eff}^{(2)}$).
The reason for this reduced effect
of the threshold shifts is that
the dominant photodisintegration effects are due anyhow to the energies around
the giant resonance peak, which for
single-nucleon emission are typically peaked around 
$ \sim  20$~MeV and for double-nucleon emission around $ \sim  26$~MeV,
and hence clearly well above the threshold energies.

\begin{figure}[t]
\centerline{{\epsfysize=3.0in \epsffile{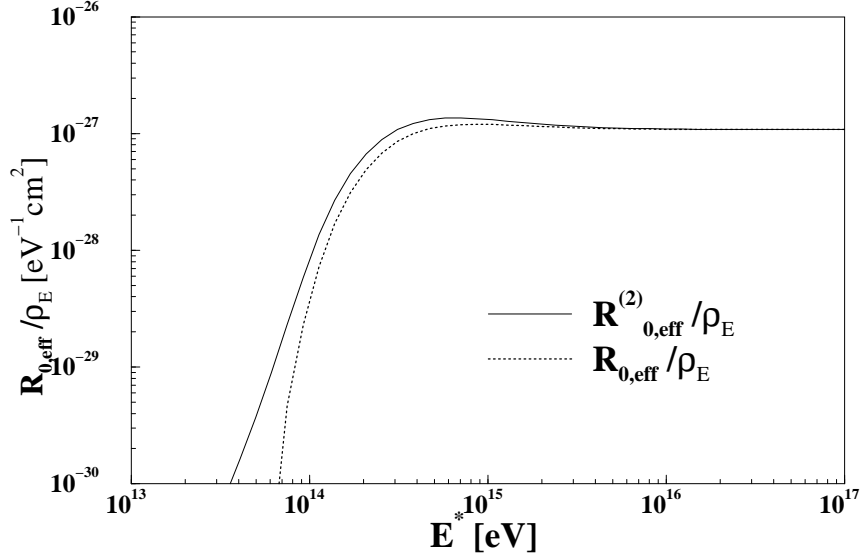}}}
\caption{Plots of $R_{0,eff}/\rho_E$ versus energy per nucleon $E^*$ 
for a thermal photon spectrum (with $k_BT=10$~eV). $R_{0,eff}^{(2)}$ 
was evaluated using
the fixed threshold energy $\epsilon_m=2$~MeV, while $R_{0,eff}$ 
corresponds to the improved threshold energies according to ref. \cite{steck1}.} 
\label{fig1}
\end{figure}

At higher energies,
$\sigma_{ij}(\epsilon')$ is approximately flat
and the multinucleon emission acquires
a higher probability, becoming actually dominant for
heavy nuclei, for which the probability for single nucleon emission is 
of only $10 \%$.
To estimate the relevance of multinucleon emission
processes, we compare the effective rate from eqs.~(6) and (7)
with the emission rate that neglects multinucleon processes
(as calculated in ref. \cite{karak}).
Figure 2(a) shows plots of $R_{i,eff}/\rho_{E}$ versus energy
per nucleon $E^*$ for a $^{56}$Fe nucleus ($i=0$) propagating through
a thermal photon spectrum (with $k_{B}T=1.8$~eV). From the figure it becomes
evident that multinucleon processes play an important role,
increasing the emission rate at high energies by a factor of $\sim 4$
with respect to the single nucleon emission results.
Analogously, Figure 2(b) shows the corresponding plots for a
power-law photon spectrum with upper energy cutoff $\epsilon_M = 20$~eV
and spectral index $\alpha = 1.3$. From this figure we can 
also see the effect
of introducing a non negligible lower cutoff in the spectrum. In fact, we observe that
with a negligible cutoff (e.g. $\epsilon_m = 10^{-6}$~eV) the emission rate
grows steadily, because increasingly abundant low energy (infrared) photons
give in this case the dominant contribution to the rates. They also
keep the influence of the giant resonance dominant even at high
particle energies, and hence in this case the multinucleon emission
has little impact.
However, if we take a non negligible cutoff into account
(for instance, $\epsilon_{m}=0.8$ eV), we find that 
the flat high energy regime of $\sigma_{i}$ now dominates beyond
a given value of $E^*$ and hence the emission rate saturates at high energies. 
Comparing the two emission rate curves with
non negligible cutoff, we observe again that multinucleon emission processes lead
to a significant increase
(by the same factor of $\sim 4$ we already encountered) in the emission rate.
Thus, introducing a non negligible lower cutoff 
energy in the power-law photon spectrum the results
should not differ much from
those with the thermal photon spectrum and, being
the rates with multinucleon emission larger, they
should also require smaller photon densities than those previously
estimated in ref. \cite {karak} to produce similar overall effects.

\begin{figure}
\centerline{{\epsfysize=3.0in \epsffile{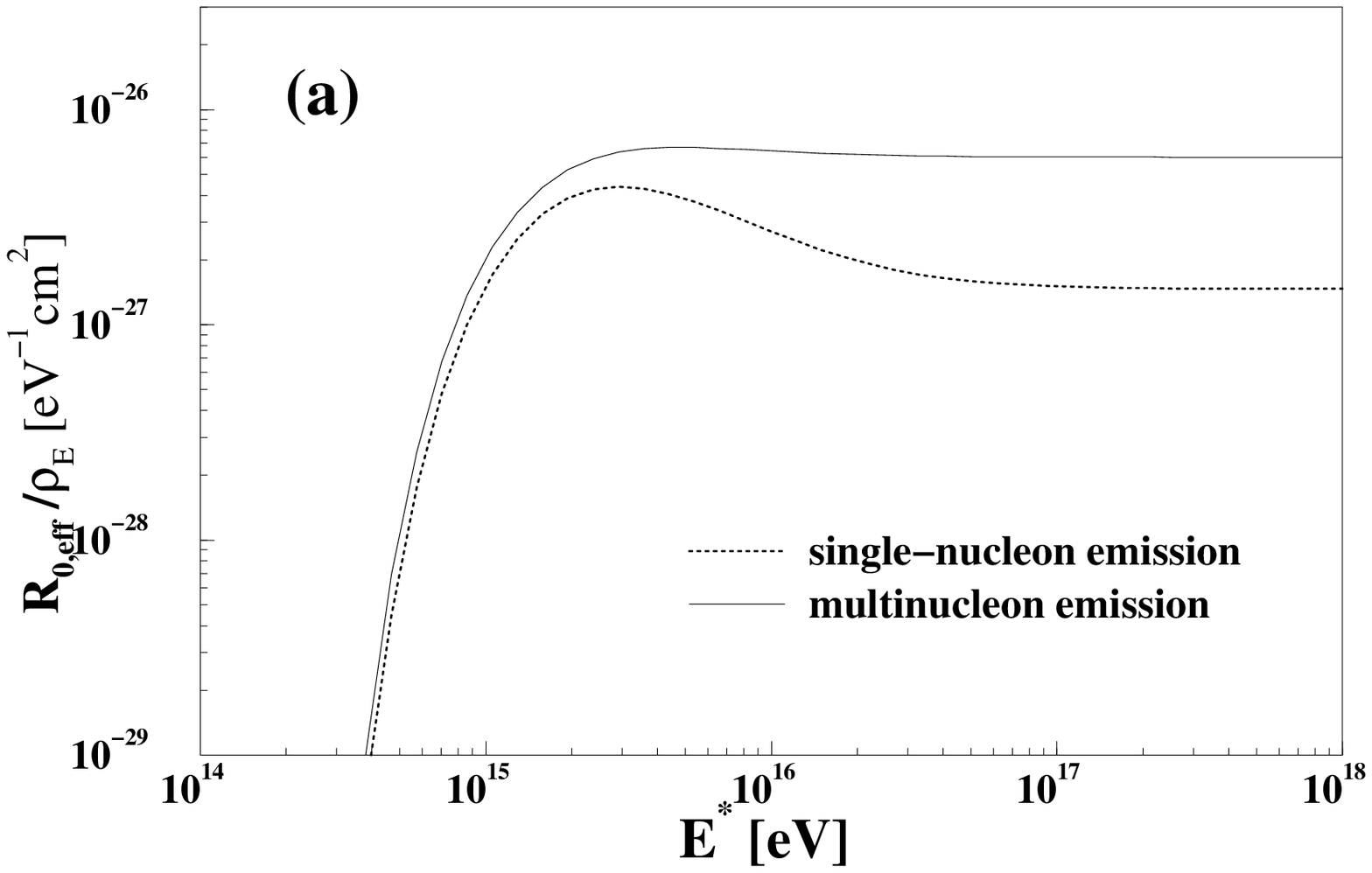}}}
\centerline{{\epsfysize=3.0in \epsffile{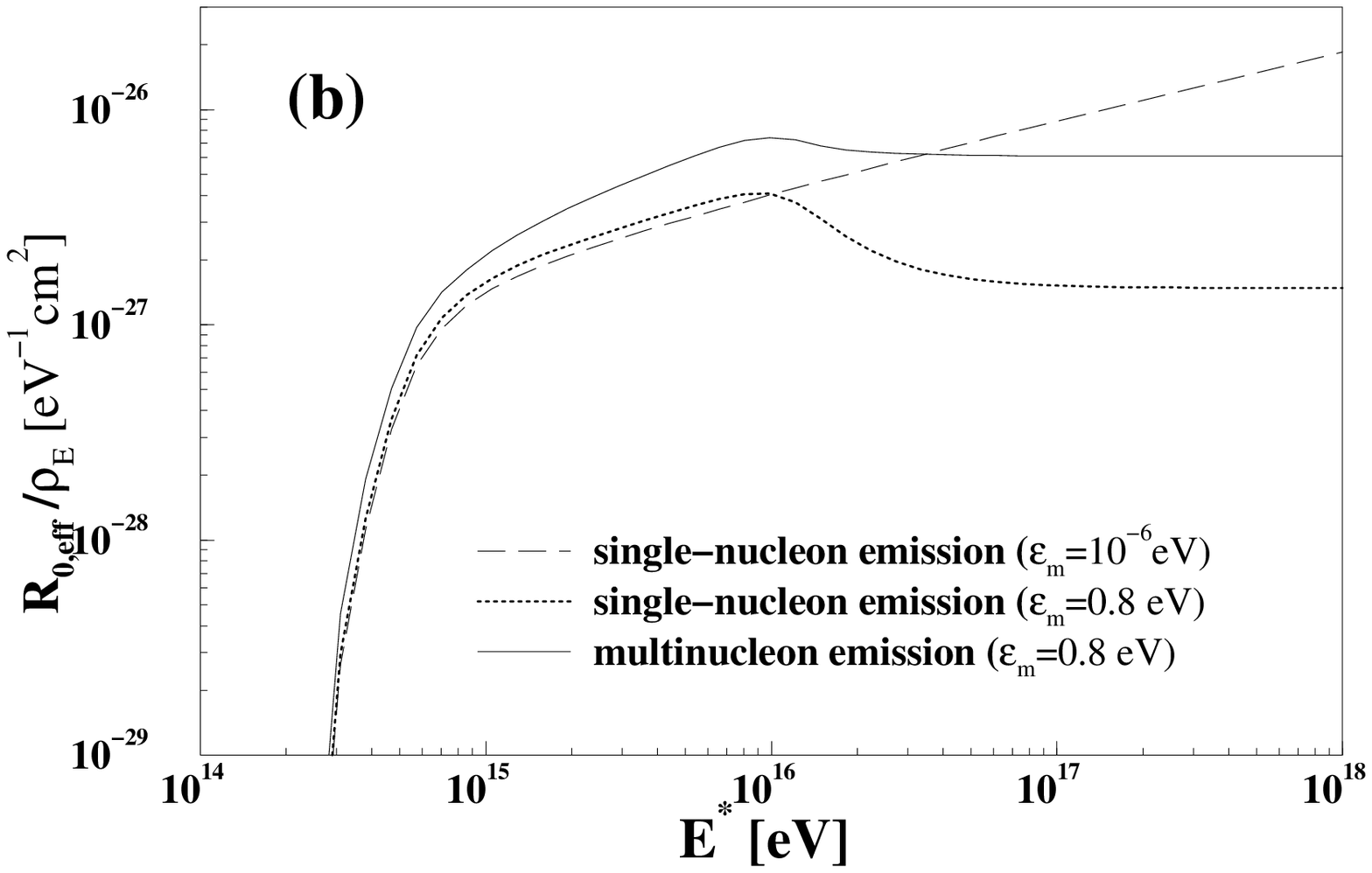}}}
\caption{Plots of $R_{0,eff}/\rho_E$ versus energy
per nucleon $E^*$. (a) Results for 
a thermal photon spectrum with $k_BT=1.8$~eV.
(b) Results for a power-law photon spectrum with upper energy cutoff 
$\epsilon_M = 20$~eV and spectral index $\alpha = 1.3$. The multinucleon
emission rate corresponds to a lower energy
cutoff $\epsilon_m = 0.8$~eV, while the single-nucleon emission rates
correspond to $\epsilon_m = 10^{-6}, 0.8$~eV, as indicated.}
\label{fig2}
\end{figure}

It should be noticed that in Figures 1 and 2 we have chosen to plot the emission
rates as functions of the energy per nucleon $E^*$,
instead of energy per nucleus $E$, 
because it is $E^*$ that remains constant during photodisintegration 
and also because for different nuclei the shapes of the cross section
are similar with the
maxima occurring
approximately at the same values of $E^*$.

The equations that describe the propagation of
nuclei are:

\begin{equation}
\frac{\partial \phi_{i}(E^{*},x)}{\partial x} = -\phi_{i}(E^{*},x)
\sum_{j\geq 1} R_{ij}(E^{*}) +
(1-\delta_{0i}) \sum_{j=1}^{i} R_{(i-j)j}(E^{*})
\phi_{i-j}(E^{*},x),
\end{equation}

\noindent where
$\phi_{i}(E^{*},x)$ is the differential flux of a nucleus
of mass $A=56-i$ (for $0 \leq i \leq 54$)
 with an energy per nucleon $E^{*}$ at
 propagation distance $x$ from the source, $R_{ij}(E^{*})$
is the above defined emission rate and $\delta_{ij}$ is
the Kronecker delta.
The exact solution of these equations (i.e., the differential
fluxes at propagation distance $L$) is given by:

\begin{equation}
\phi_{i}(E^{*},L) = \sum_{j=0}^{i} b_{ij}(E^{*})
\exp (-\sum_{k\geq 1} R_{jk}(E^{*}) L) ,
\end{equation}

\noindent where

\begin{equation}
b_{ii}=\phi_i^0 (E^{*})-(1-\delta_{0i})\sum_{j=0}^{i-1} b_{ij}(E^{*})
\end{equation}

\noindent and 

\begin{equation}
b_{ij} = {\sum_{k=1}^{i-j} b_{(i-k)j}(E^{*}) R_{(i-k)k}(E^{*})\over 
 \sum_{k \geq 1} R_{ik}(E^{*})-\sum_{k \geq 1} R_{jk}(E^{*})},\ \ 
 {\rm for}\ \ i>j.
\end{equation}

Recalling that the emission rates $R_{ij}$ are linear in the photon
energy density $\rho_E$, it should be noted that, as expected, the fluxes
only depend on the column density $\rho_E L$. It also should be pointed out
that, since the equations describing the CR propagation 
only depend on the energies per
nucleon $E^{*}$, the final CR spectra 
require the re-calculation of the solution (eqs.(9)-(11)) in terms of
energy per nucleus. 

\subsection{ Propagation of protons}

The main energy loss mechanism that must
be taken into account for the propagation of protons 
is the photomeson production
process.  
We have to distinguish between the primary protons
produced at the source with an initial flux $\phi_{i=55}^0$
(as given by eq.(1)) and the nucleons released as by-products
of the nuclear photodisintegration. The fluxes corresponding
to these (secondary) protons and neutrons
($\phi^{sp}$ and $\phi^n$, respectively) can be easily
determined from the solution for the nuclear fluxes, as we will
see in the following.
Let $\phi_i^N (E,x)$ be the differential flux of nucleons of energy $E$
produced within a distance $x$ away from the source by the
photodisintegration of nuclei of mass number $A=56-i$ and energy
per nucleon $E^{*}=E/A$. Then, the equation governing the 
evolution of the fluxes is

\begin{equation}
\frac{\partial \phi_i^N(E,x)}{\partial x} =
\phi_{i}(E^{*},x) \sum_{j \geq 1} j R_{i,j}(E^{*})
\ \ , \ \ {\rm for}\  0 \leq i \leq 54 \ \ ,
\end{equation}

\noindent (with the initial condition $\phi_i^N (E,x=0) = 0$)
and its solution reads

\begin{equation}
\phi_i^N (E,L) = \sum_{j=0}^{i}
\left(\frac{\sum_{k \geq 1} k R_{ik}(E^{*})}{\sum_{k \geq 1}
R_{jk}(E^{*})}\right) 
b_{ij}(E^{*})
\left[ 1-\exp \left(-\sum_{k\geq 1} R_{jk}(E^{*}) L\right) \right] \ \ .
\end{equation}

Thus, the total fluxes of secondary protons and neutrons are
found by performing the weighted sums over index $i$:

\begin{equation}
\phi^{sp}(E,L) = \sum_{i=0}^{54} \frac{Z_i}{56-i} \phi_i^N (E,L) \ \ ,
\end{equation}
\begin{equation}
\phi^{n}(E,L) = \sum_{i=0}^{54} \left(1 -\frac{Z_i}{56-i}\right)
 \phi_i^N(E,L) \ \ ,
\end{equation}
where $Z_{i}$ is the charge of the $i$-nucleus.

We will also discuss the possible impact of the so-called
neutron mechanism\cite{neutron}, 
which could take place in the presence of 
a magnetic field in the source region. Indeed, a magnetic field
would retain the charged particles within
that region for some period of time, while non-charged particles
would be able to escape rapidly.
In that case, we will consider that the flux
of neutrons $\phi^n$ adds to the final all-particle CR flux
neglecting any energy losses, 
while the flux of secondary protons $\phi^{sp}$, combined with
the initial flux of primary protons $\phi_{i=55}^0$,
will undergo energy losses due to the well studied
process of photopion production. Indeed, since a background
photon looks like a high-energy gamma ray in the proton
rest frame, the photopion reactions

\begin{equation}
p + \gamma \to p + \pi^{0}
\end{equation}
\begin{equation}
p + \gamma \to n + \pi^{+}
\end{equation}
proceed with an interaction probability given by

\begin{equation}
g(E)={1\over 2\gamma^2}\int_{\epsilon'_{thr}/2\gamma}^\infty
{\rm d}\epsilon
{n(\epsilon)\over \epsilon^{2}}\int_{\epsilon'_{thr}}^{2\gamma\epsilon}
{\rm d}\epsilon'\epsilon'\sigma_{\gamma p}(\epsilon')K(\epsilon'),
\end{equation}

\noindent where, analogously to eq.~(6), $\gamma=E/m_pc^2$ is 
the Lorentz factor
of the proton, $\sigma_{\gamma p}$ is the $\gamma p$ interaction cross
section (parametrized from experimental data compiled
in ref.\cite {table}), $K$ is the coefficient of inelasticity
(defined as the average relative energy loss of the proton),
$\epsilon$ the photon energy in the observer's system and
$\epsilon'$ its energy in the rest frame of the proton. The
energy threshold for photopion production is $\epsilon'_{thr}=145$ MeV.

Then, if we call $\phi^{p}(E,x)$ and $\phi^{*n}(E,x)$
the differential fluxes of protons and neutrons of energy $E$ (produced
by the reactions given by eqs.(16) and (17), respectively)
after tarversing a distance $x$ from the source, the corresponding
propagation equations are

\begin{equation}
\frac{\partial \phi^{p}(E,x)}{\partial x} =
-g(E) \phi^{p}(E,x) + \frac{1}{2}\frac{1}{(1-K)} g\left(\frac{E}{1-K}\right)
\phi^{p}\left(\frac{E}{1-K},x\right)
\end{equation}

\noindent and (neglecting neutron energy losses)

\begin{equation}
\frac{\partial \phi^{*n}(E,x)}{\partial x} =
\frac{1}{2}\frac{1}{(1-K)} g\left(\frac{E}{1-K}\right)
\phi^{p}\left(\frac{E}{1-K},x\right) \ \ ,
\end{equation}

\noindent where the factors $\frac{1}{2}$ arise from the relative
probability of occurrence of reactions (16) and (17).
The coefficient of inelasticity $K$ was given a typical
value $K=0.3$, irrespective of proton energy \cite{karak}.
The initial conditions to solve these differential equations are

\begin{equation}
\phi^{p}(E,x=0) = \phi_{i=55}^0(E) + \phi^{sp}(E,L)
\end{equation}
and

\begin{equation}
\phi^{*n}(E,x=0) = 0 \ \ ,
\end{equation}
respectively. In the first one we assumed for simplicity
that the secondary protons produced by photodisintegration of
nuclei add directly to the initial proton spectrum. 
This is justified since whenever the photopion processes are
relevant, the photodisintegration is very efficient
and hence nuclei are rapidly disintegrated. 
The solution to eq.(19) turns out to be

\begin{equation}
\phi^{p}(E,L) = \sum_{l=0}^{\infty} \phi_l^p(E,L) \ \ , 
\end{equation}

\noindent where

\begin{equation}
\phi_0^p(E,L) = \phi^{p}(E,0) \exp(-g(E) L) \ \ ,
\end{equation}
while for $l>0$
\begin{equation}
\phi_l^p(E,L)= {\phi^{p}\left(\frac{E}{(1-K)^{l}},0\right)\over (2(1-K))^{l}}
\prod_{j=1}^{l} g\left(\frac{E}{(1-K)^{j}}\right)
\sum_{n=0}^{l} {\exp \left(-g\left(\frac{E}{(1-K)^{n}}\right) L\right)\over
\prod_{^{m=0} _{m \neq n}}^{l}
\left[
g\left(\frac{E}{(1-K)^{m}}\right)-g\left(\frac{E}{(1-K)^{n}}
\right)\right]} \ \ .
\end{equation}

For the neutron component, the solution to eq.(20) reads

\begin{equation}
\phi^{*n}(E,x) = \sum_{l=0}^{\infty} \phi_l^{*n}(E,x) \ \ , 
\end{equation}
where

\begin{equation}
\phi_0^{*n}(E,L)= \frac {\phi^{p}\left(\frac{E}{1-K},0\right)}{2(1-K)}
\left(1-\exp\left(-g\left(\frac{E}{1-K}\right) L\right)\right) \ \ ,
\end{equation}
while for $l> 0$ 
$$ 
\phi_l^{*n}(E,L)= \frac {\phi^{p}\left(\frac{E}{(1-K)^{l+1}},0\right)}
{(2(1-K))^{l+1}}
\prod_{j=0}^{l} g\left(\frac{E}{(1-K)^{j+1}}\right)\times
$$
\begin{equation}
\sum_{n=0}^{l} {1-\exp\left(-g\left(\frac{E}{(1-K)^{n+1}}\right) L\right) 
\over g\left(\frac{E}{(1-K)^{n+1}}\right)
\prod_{^{m=0}_{ m \neq n}}^l
\left[g\left(\frac{E}{(1-K)^{m+1}}\right)-g\left(\frac{E}{(1-K)^{n+1}}
\right)\right]}.
\end{equation}

The linearity of the photopion reaction rate $g(E)$ with respect to
the photon energy density $\rho_E$ (as implied by eq.(18))
causes the proton and neutron fluxes to depend only on the
column density $\rho_E L$, similarly as for the
fluxes of nuclei. \\

\section{Results and comparison with observations}

So far, we wrote down the appropriate equations governing
the propagation of CR particles, and we found their exact solutions:
eqs.(9)-(11) for nuclei, eqs.(13),(15) and (26)-(28) for neutrons (that
eventually decay into protons outside the source region), and eqs.(23)-(25)
for protons (that undergo energy losses due to photopion production),
the sum of all these contributions being the total CR flux $\phi_{total}$.
We can now investigate how $\phi_{total}$ depends on
the assumed photon distributions around 
the source, namely on the column density $\rho_E L$ and on
the parameters involved in the photon spectral distributions (eqs.(2) and (3)).

Figure 3(a) shows the total CR differential flux as a function of
the energy per particle for a Planckian spectrum with $k_{B}T=1.8$ eV
and several values for the column density $\rho_E L$. Analogously,
Figure 3(b) shows the results corresponding to
a Planckian spectrum with $k_{B}T=10$ eV. To compare our results
with experimental data, the figures also show the observed spectra
measured by different experiments (CASABLANCA, DICE, Tibet, PROTON
satellite and AKENO) \cite{fowler,swordy,amen2}.

\begin{figure}
\centerline{{\epsfysize=3.0in \epsffile{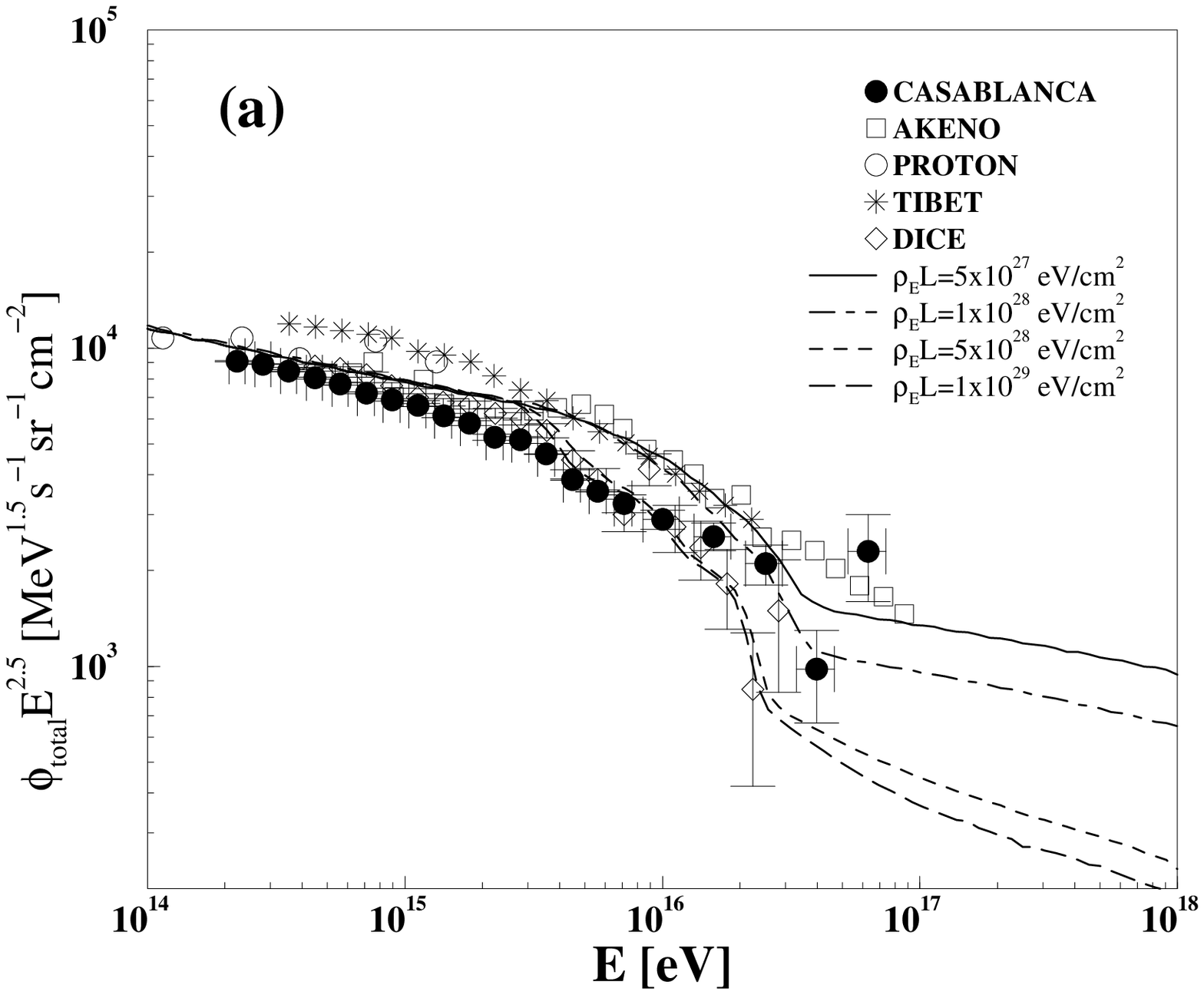}}}
\centerline{{\epsfysize=3.0in \epsffile{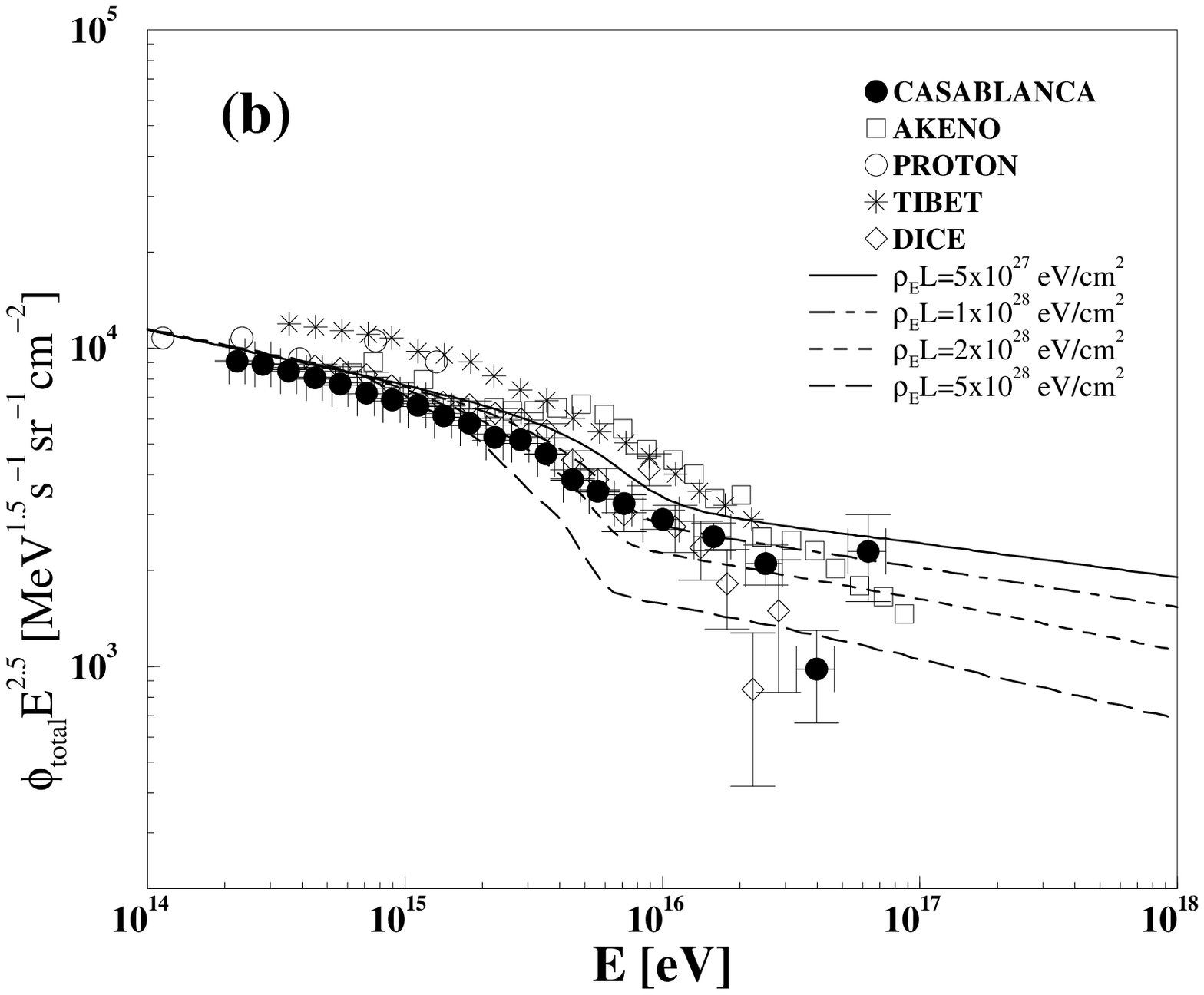}}}
\caption{Plots showing the total CR flux versus the energy
per particle for thermal spectra with (a) $k_BT=1.8$~eV and (b) $k_BT=10$~eV,
and several values for the column density $\rho_EL$, as indicated.
For the sake of comparison, experimental data measured by different
experiments are also shown.}
\label{fig3}
\end{figure}
 
The effect of increasing the temperature is clearly that of 
shifting the knee towards lower energies,
while the effect of increasing $\rho_E L$ is that of intensifying the
steepening at the knee. This behavior is easily explained by the fact
that it is the nuclear photodisintegration the responsible for the
occurrence of the knee (since photopion production processes
show up at higher energies). Then, if the photon temperature
(and thus, the mean photon energy) is larger, the particle energy required
for the nuclei to disintegrate decreases.
Furthermore, CR propagation cannot be affected by $\rho_E L$
below the knee, where the photodisintegration has not set
up yet, but above the knee 
the disintegration rates have a linear dependence on $\rho_E$.
Hence, we should expect $\rho_E L$
to control the steepening in the knee region, in agreement
with the results shown in the figures.

In ref. \cite {karak} the best fit to
observations for the Planckian spectrum was
that corresponding to $k_{B}T=1.8$~eV and $\rho_E
L=2.25\times10^{29}$~eV/cm$^2$. 
For the sake of comparison,  
Figure 3(a) shows our results for the same temperature.
It is clearly observed that the
new results are compatible with the AKENO experimental data for 
low column density values 
$\rho_E L\simeq 5\times 10^{27}$
~eV/cm$^2$, i.e.  $\sim 50$ times lower than the previous result.
Figure 3(b) shows that the corresponding results for $k_{B}T=10$~eV
are in somewhat better agreement with CASABLANCA observations for 
column densities in the range
$\rho_E L=5\times 10^{27}-2\times 10^{28}$~eV/cm$^2$. 
One has to keep in mind however that the model discussed 
is oversimplified in many respects: we assume that the same photon column 
density is traversed by all nuclear species and for all CR energies, 
we neglect additional effects due to change in diffusion 
properties at these energies, which in alternative scenarios \cite{ptus}
are actually the responsible for the appearance of the knee, we assume
that CR interactions in the interstellar medium are negligible, etc..
Hence, an accurate agreement  with observational data is not to be 
expected, although it is reassuring to see that the results have 
the right tendency to account for the main qualitative features 
of the data in the knee region.

Figures 4(a) and 4(b) show plots of mean mass composition 
$\langle \rm{ln} A \rangle$ versus $E$
corresponding to the Planckian photon spectra with $k_BT=1.8$~eV and
$k_BT=10$~eV, respectively.
For comparison, we also show the CR mass composition measured by
different experiments \cite{fowler,swordy,ka01,chat}.
One can observe that solutions corresponding to
the largest plotted values of column density $\rho_E L$
imply that CRs above the knee are chiefly
constituted by protons, in disagreement with experimental data; the
non-negligible contribution of heavier CR components beyond the knee
imposes an upper limit on the column density around $(\rho_EL)_{max} \approx
2 \times 10^{28}$~eV/cm$^2$.
This result
stresses again the fact that multinucleon emission processes play
a significant role in CR propagation within this scenario.
We also see that, although the solutions with lower column density
($\rho_EL \leq (\rho_EL)_{max}$) are in reasonable agreement with
 some data sets
within error bars, the composition beyond the knee seems to be
somewhat suppressed. Nevertheless, we should bear in mind that we have
employed just one source with a fixed column density,
and that in a realistic situation
there would be many sources, each with
a different environment of radiation.
Avoiding any detailed calculations, 
we may think about the contribution of one additional source.
If it happens to have a column density $\rho_E L$ somewhat smaller,
it would lead to a flatter flux; then, whereas the
first source would be responsible for the occurrence of the knee,
the second one would keep a non-vanishing nuclear component beyond the knee.
Furthermore, in the computations we assumed that all nuclei traverse the same
distance $L$, thus disregarding the fact that the magnetic
field present in the source region should retain more effectively the heavier
nuclei at any given CR energy. Taking this into account (for instance, assuming that a nucleus traverses
a distance in the photon background proportional to its charge) we would find that, while the
disintegration would be very effective for heavy nuclei
(and would be, in fact, responsible for forming the knee), it would be
less effective for the lighter nuclei (since the column density for them
would be much smaller) and they could then survive beyond the knee.
Hence, we see that this scenario can easily account for a change in the
composition towards lighter nuclei above the knee, 
which is apparent in some of the existing data. On the contrary,
rigidity dependent scenarios predict the opposite trend, and
may have difficulties to account for some measurements. It has indeed
been pointed out that they would require the introduction of 
an ad-hoc additional light component above the knee to reproduce
the observations \cite{arm}.

\begin{figure}
\centerline{{\epsfysize=3.0in \epsffile{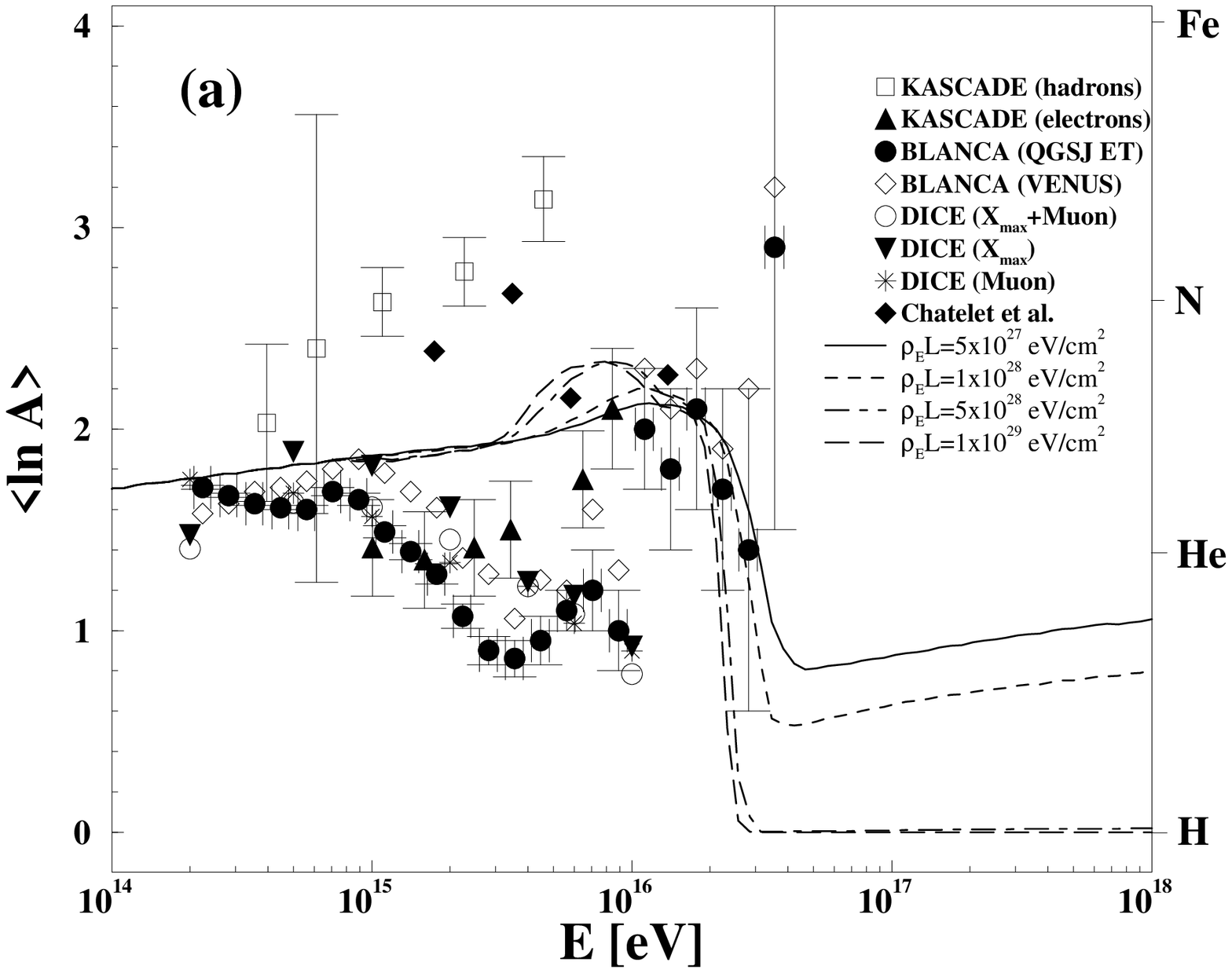}}}
\centerline{{\epsfysize=3.0in \epsffile{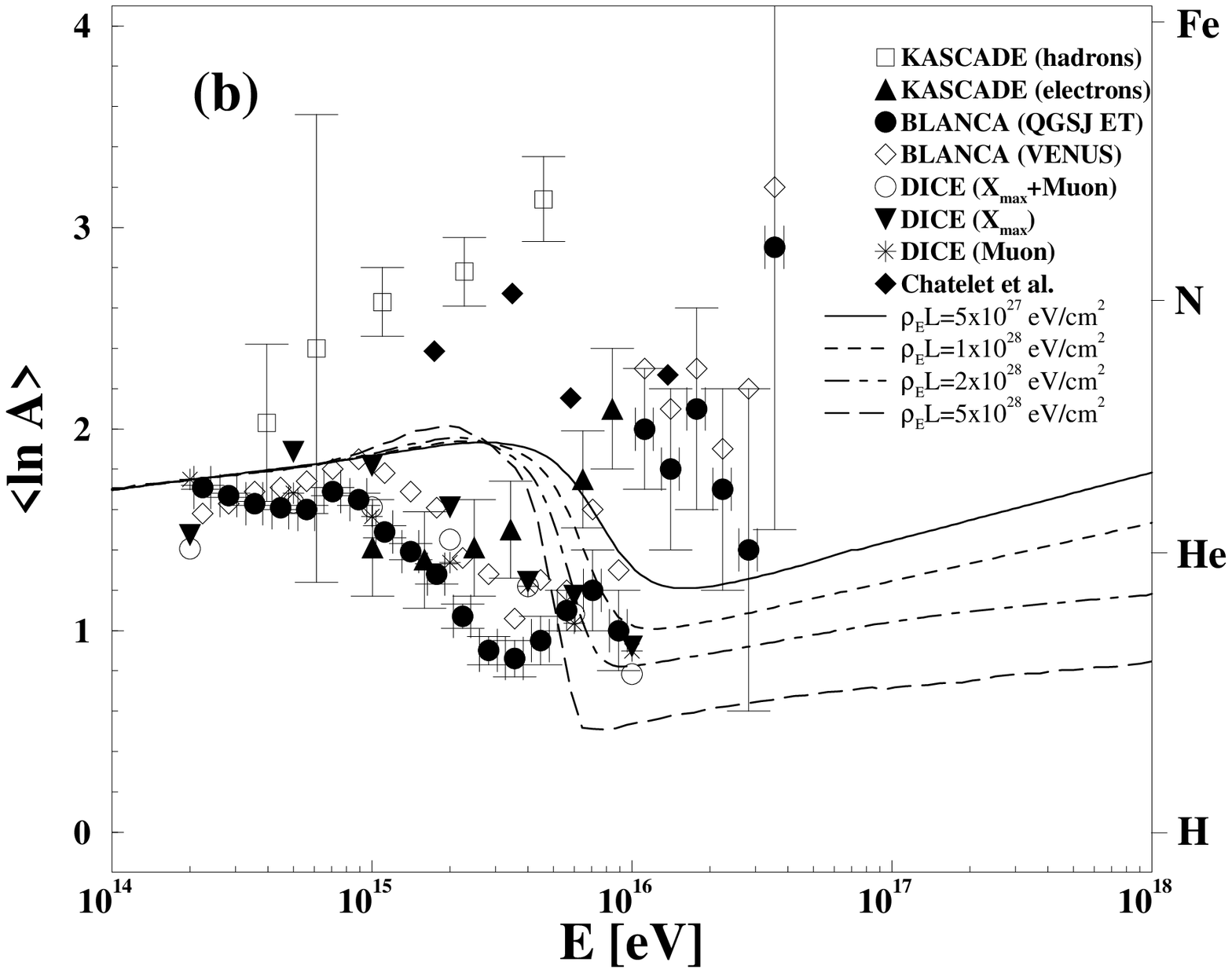}}}
\caption{Plots of mean mass composition $\langle \rm{ln} A \rangle$ versus $E$
corresponding to the Planckian photon spectra with (a) $k_BT=1.8$~eV and
(b) $k_BT=10$~eV. For comparison, the CR mass composition measured by
different experiments is also shown.
Notice that several data sets exhibit the same
qualitative behavior, since they correspond to the same experimental
data and differ only in the hadronic models used to interpret the
observations.}
\label{fig4}
\end{figure}

The results for the power-law photon spectrum (with
a lower cutoff energy in the infrared range) 
should not differ much from the ones
obtained for the Planckian spectrum, and
this behavior was in fact corroborated. 
Let us fix the spectral index and the upper cutoff energy at
the values found in ref. \cite{karak} to correspond
to the best fit to experimental data, namely $\alpha=1.3$ and
$\epsilon_M=20$~eV. 
Figure 5 shows $\phi_{total}$ versus $E$ for power-law distributions
with column density $\rho_E L=10^{28}$~eV/cm$^{2}$
and several values for the lower cutoff energy
($\epsilon_m=0.01,0.1,0.8$~eV). It is clear that the 
overall spectrum  displays a steepening which has the 
right qualitative features to explain the knee although 
it does not provide an accurate fit to neither AKENO nor 
CASABLANCA data individually. 
Turning now to
discuss the dependence of the results on the lower cutoff energy,
it is found that, apart from minor differences arising
from spectrum normalization,
lower values of $\epsilon_m$ yield lower fluxes in the high energy end.
The explanation for this is that from
the expression for the photodisintegration emission rate (eq.(6))
it turns out that the role of $\epsilon_m$ is that of
determining a value of the particle energy $E$, such that
below $E$ the giant dipole resonance dominates, while above $E$
the flat high energy regime of $\sigma$ prevails. As we lower
$\epsilon_m$, more photons with lower energy become available so that
the flux suppression imposed by the giant resonance extends
to higher particle energies.

\begin{figure}[t]
\centerline{{\epsfysize=3.0in \epsffile{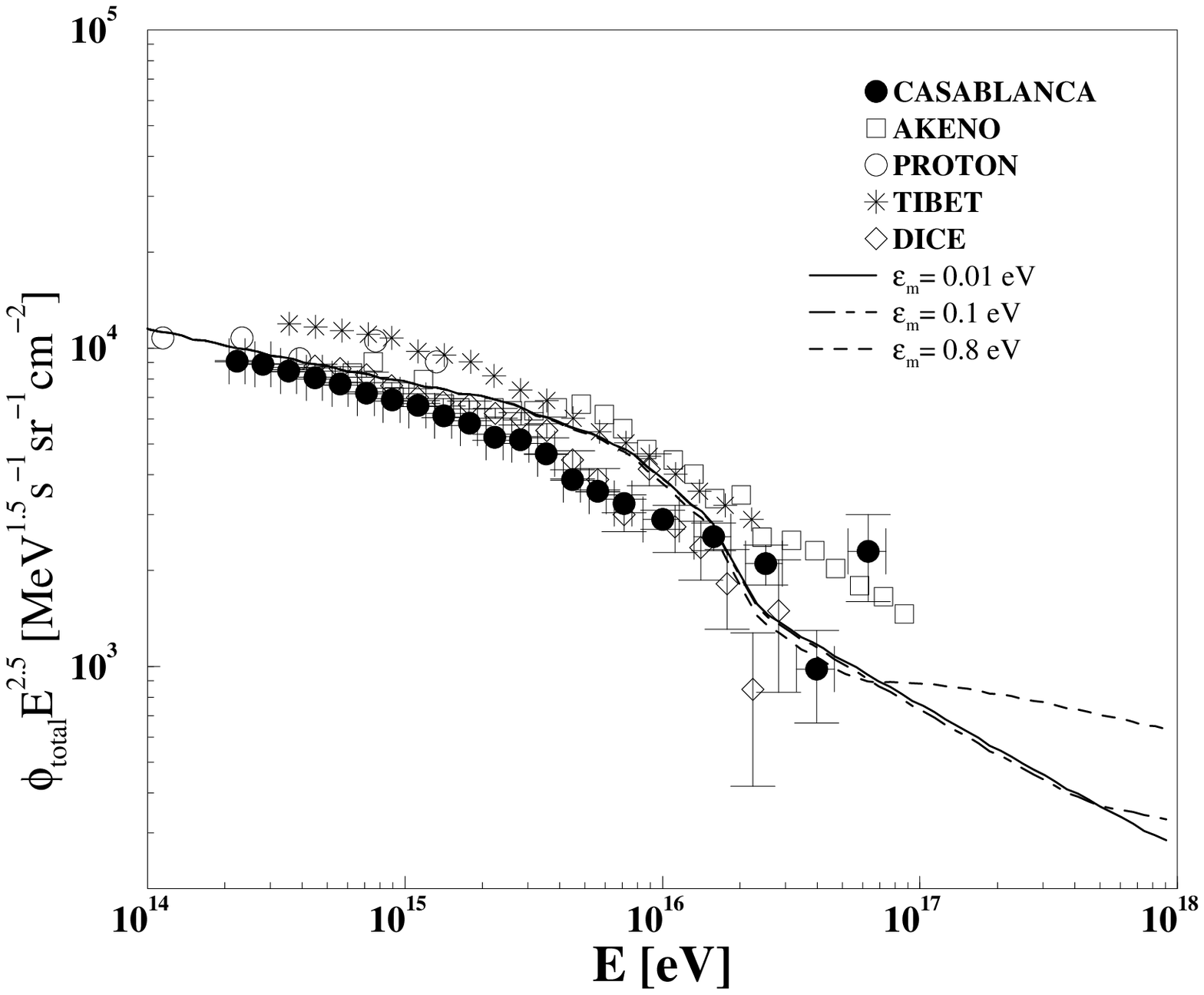}}}
\caption{Plots showing the total CR flux versus the energy
per particle for power-law spectra with
$\alpha=1.3$, $\epsilon_M=20$~eV, $\rho_E L=10^{28}$~eV/cm$^{2}$ and 
several values for the lower cutoff energy
($\epsilon_m=0.01,0.1,0.8$~eV).
For comparison, experimental data measured by different
experiments are also shown.}
\label{fig5}
\end{figure}

This analysis provides a suitable explanation of the results
shown in Figure 5, as well as those of Figure 2(b).
The results previously obtained for power-law spectra \cite{karak}
show an extremely abrupt flux suppression that seems to be in contrast
with experimental data (CR energy spectra and, especially, CR mass
composition) above $E=10^{17}$~eV, and can be attributed to the
negligible value for $\epsilon_m$ adopted there. The use of a
non negligible cutoff
prevents the flux suppression and provides a better fit to observations in
this energy region.
It should also be remarked that a solution yielding an excessive
CR flux at energies above $10^{17}$ eV
is by no means troublesome since it is clear
that at these high energies efficient leakage of CRs from the Galaxy
cannot be disregarded, and that a less effective acceleration
could also be playing a role. 
These mechanisms would clearly contribute to lower the CR flux, so that it does not
seem reasonable to expect a reliable explanation of the CR spectrum
by photodisintegration processes alone beyond $E\sim 10^{17}$~eV.

\begin{figure}[t]
\centerline{{\epsfysize=3.0in \epsffile{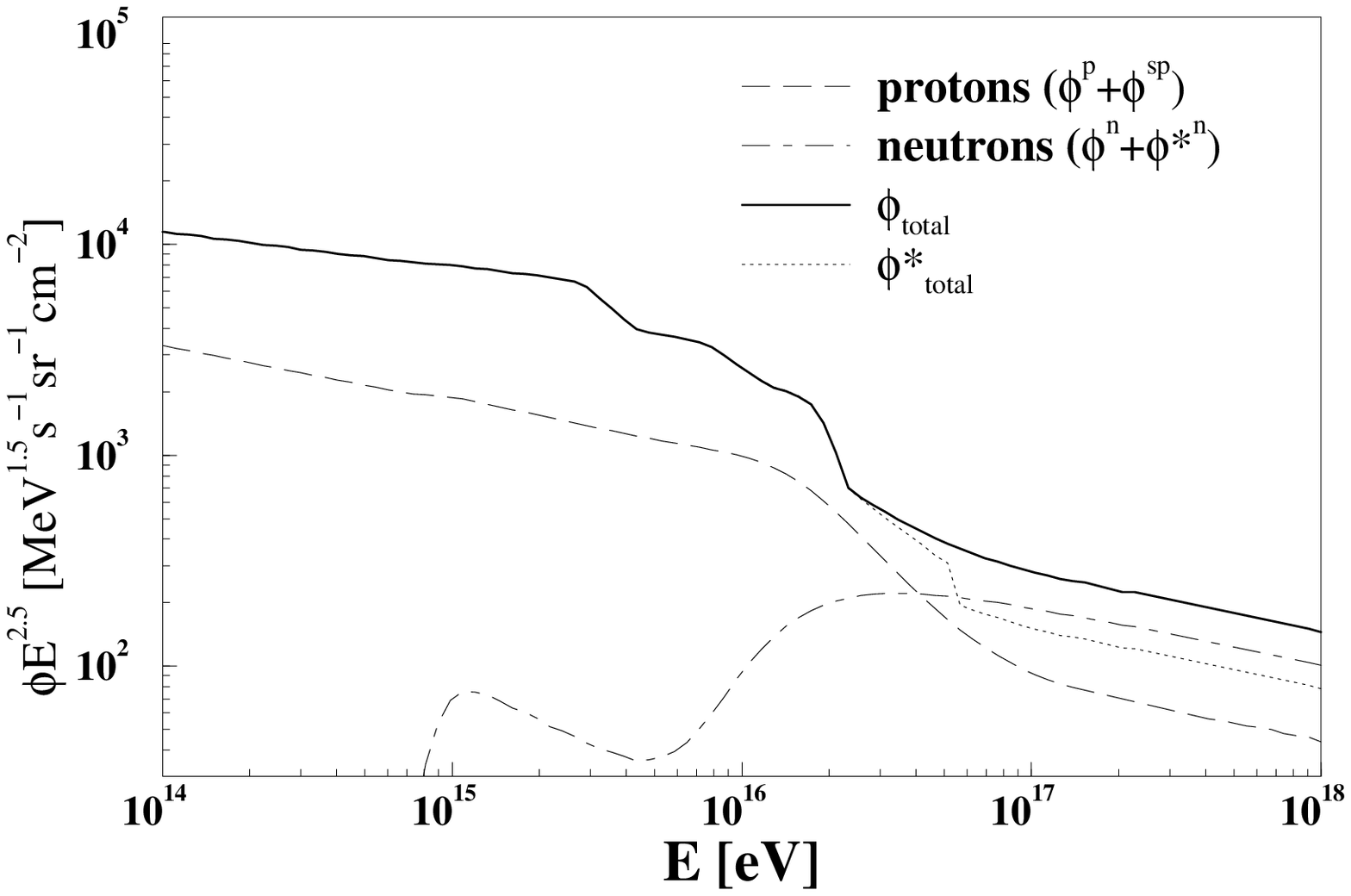}}}
\caption{Plots showing the proton flux
($\phi^p + \phi^{sp}$), the neutron flux ($\phi^n + \phi^{*n}$)
and the total CR flux ($\phi_{total}$) versus the energy
per particle for a thermal photon spectrum with $k_BT=1.8$~eV
and $\rho_E L=2 \times 10^{29}$~eV/cm$^2$.
Also shown is the total CR flux ($\phi_{total}^*$) 
that is obtained ignoring the escape of neutrons.}
\label{fig6}
\end{figure}

Figure 6 shows plots of the proton and neutron components ($\phi^p+
\phi^{sp}$ and $\phi^n + \phi^{*n}$, respectively) as well as the
total CR flux
$\phi_{total}$ for the Planckian photon spectrum with $k_BT=1.8$~eV
and $\rho_E L=2 \times 10^{29}$~eV/cm$^2$. In order to estimate the possible effects of
the neutron mechanism, we also plotted the total CR flux
that is obtained ignoring the escape of neutrons, labelled 
$\phi_{total}^*$. From the figure it is seen that the 
neutron mechanism raises the
CR total flux by $\sim 100 \%$ for $E \geq 5 \times 10^{16}$~eV. Had we
adopted a column density $\rho_E L=10^{29}$~eV/cm$^2$, the corresponding 
flux increment would be only $\sim 10 \%$. Nevertheless,
it should be noted that these results correspond to very high column
densities (actually inconsistent with the observed CR composition), 
and that effects arising from the neutron escape mechanism
turn out to be negligible for lower values of the column density $\rho_EL$,
for which also the photopion losses of protons can be altogether ignored.

To see what are the implications for the sources in order to produce the
 required photon densities for this scenario to work, let us make 
the following simple estimate. An isotropic source with photon energy 
density $\rho_{E,0}$ at its surface (with radius $r_0$) will have around 
it a photon energy density given by $\rho_E(r)=\rho_{E,0}(r_0/r)^2$. 
This means 
that the photon 
column density measured radially from the surface is $\langle \rho_E r\rangle\simeq \rho_{E,0} r_0$. On the other 
hand, if the CRs are magnetically confined near the source, the path 
they travel across the dense photon field will be larger, i.e. 
$\langle \rho_E L\rangle\equiv \lambda \langle \rho_E r\rangle$, with 
the parameter $\lambda\gg 1$.  If to obtain an estimate of the maximum
achievable densities we assume that the source radiates with a 
luminosity $L=4\pi r_0^2\rho_{E,0} c$ smaller than the Eddington 
limit, i.e. $L<L_{Edd}=4\pi GMm_pc/\sigma_T\simeq 1.3\times 
10^{38}M/M_\odot$~erg/s, with $M$ the mass of 
the source, $m_p$ the proton mass and $\sigma_T$ Thomson's cross 
section, we find that the radius of the source needs to satisfy
\begin{equation}
r_0<\frac{GMm_p}{\sigma_T\rho_{E,0}r_0}\simeq \left(\frac{\lambda}
{10}\right)\left(\frac{10^{28}{\rm eV/cm}^2}{\langle \rho_E L\rangle}\right)
\left( \frac{M}{M_\odot}\right)2\times 10^9 {\rm m},
\end{equation}
and hence the photon source must be compact (like a pulsar) and 
in general smaller than the typical size of a white dwarf. 
It has 
to be mentioned however that in non-steady-state situations like 
in supernova explosions, luminosities larger than $L_{Edd}$ are possible.
Another consideration is that if the source has a black-body spectrum 
with temperature $T$ and it is radiating at the Eddington limit, the
relation $4\pi r_0^2\sigma T^4=L_{Edd}$ implies that
$T\simeq 1.8\times 10^3$ eV$ (M/M_\odot)^{1/4}(10$~km/$r_0)^{1/2}$, 
and hence for the photon energies to be in the optical/UV region, 
radii somewhat larger than those of neutron stars would be favored (or the
luminosity should be smaller than $L_{Edd}$).

As a summary, we have considered in more detail the photodisintegration
of CR nuclei in the scenario in which their interaction with optical and
soft UV photons around the source is responsible for the steepening of 
the spectrum and for the change in composition above the knee.
The more accurate treatment of the nuclear processes, in particular
the inclusion of the multinucleon emission, implies that the 
required photon column densities are lower (by an order of magnitude)
than previously obtained. 
The main difference between this scenario and other (rigidity dependent)
explanations proposed for the knee is the resulting CR composition,
which here becomes lighter above $\sim 10^{16}$~eV.  

\section*{Acknowledgments}
Work supported by CONICET, ANPCyT and Fundaci\'on Antorchas, 
Argentina. E. R. thanks A. Dar for fruitful discussions on the subject.

\end{document}